\documentclass[conference]{IEEEtran}
\IEEEoverridecommandlockouts
\usepackage{cite}
\usepackage{amsmath,amssymb,amsfonts}
\usepackage{algorithmic}
\usepackage{graphicx}
\usepackage{textcomp}
\usepackage{xcolor}
\usepackage{multirow}

\usepackage[switch, modulo]{lineno}
\def\BibTeX{{\rm B\kern-.05em{\sc i\kern-.025em b}\kern-.08em
    T\kern-.1667em\lower.7ex\hbox{E}\kern-.125emX}}
\begin{document}

\title{Design and Assessment of a Bimanual Haptic Epidural Needle Insertion Simulator \\

\author{Nitsan Davidor, \textit{Student Member, IEEE}, Yair Binyamin, Tamar Hayuni and Ilana Nisky, \textit{Senior Member, IEEE}}

\thanks{This research was supported in part by the Israeli Science Foundation (grant 327/20) and by the Helmsley Charitable Trust through the Agricultural, Biological and Cognitive Robotics Initiative and by the Marcus Endowment Fund, both at Ben-Gurion University of the Negev.}
\thanks{Nitsan Davidor and Ilana Nisky are from the Department of Biomedical Engineering and the School of Brain Sciences and Cognition, Ben-Gurion University of the Negev, Beer Sheva, Israel.
        {\tt\small nitsanti@post.bgu.ac.il,
        nisky@bgu.ac.il}}%
\thanks{Tamar Hayuni and Yair Binyamin are from the Department of Anesthesiology, Soroka Medical Center, Ben-Gurion University of the Negev, Beer Sheva, Israel.
        {\tt\small yairben1@gmail.com, 
        tamarhayuni@gmail.com}}%
}

%\author{\IEEEauthorblockN{1\textsuperscript{st} Nitsan Davidor}
%\IEEEauthorblockA{\textit{Biomedical Engineering} \\
%\textit{Ben Gurion University of the Negev}\\
%Be'er Sheva, Israel \\
%nitsanti@post.bgu.ac.il}
%\and
%\IEEEauthorblockN{2\textsuperscript{nd} Tamar Hayuni}
%\IEEEauthorblockA{\textit{Department of Anesthesiology} \\
%\textit{Soroka Medical Center}\\
%Be'er Sheva, Israel \\
%tamarhayuni@gmail.com}
%\IEEEauthorblockN{3\textsuperscript{rd} Yair Binyamin}
%\IEEEauthorblockA{\textit{Department of Anesthesiology} \\
%\textit{Soroka Medical Center}\\
%Be'er Sheva, Israel \\
%yairben1@gmail.com}
%\and
%\IEEEauthorblockN{4\textsuperscript{th} Ilana Niksy}
%\IEEEauthorblockA{\textit{Biomedical Engineering} \\
%\textit{Ben Gurion University of the Negev}\\
%Be'er Sheva, Israel \\
%nisky@bgu.ac.il}
%}

\maketitle

\begin{abstract} %200 words
    The case experience of anesthesiologists is one of the leading causes of accidental dural punctures and failed epidurals - the most common complications of epidural analgesia used for pain relief during delivery. We designed a bimanual haptic simulator to train anesthesiologists and optimize epidural analgesia skill acquisition. We present an assessment study conducted with 22 anesthesiologists of different competency levels from several Israeli hospitals. Our simulator emulates the forces applied to the epidural (Touhy) needle, held by one hand, and those applied to the Loss of Resistance (LOR) syringe, held by the other one. The resistance is calculated based on a model of the epidural region layers parameterized by the weight of the patient. We measured the movements of both haptic devices and quantified the results' rate (success, failed epidurals, and dural punctures), insertion strategies, and the participants' answers to questionnaires about their perception of the simulation realism. We demonstrated good construct validity by showing that the simulator can distinguish between real-life novices and experts. Face and content validity were examined by studying users' impressions regarding the simulator's realism and fulfillment of purpose. We found differences in strategies between different level anesthesiologists, and suggest trainee-based instruction in advanced training stages. 
    
\end{abstract}

\section{Introduction}
Delivery is one of the most painful experiences in a woman's life, and epidural analgesia is the most common option for pain relief during delivery with a rate of 60\% in Israel \cite{10year} and 
71\% in the United States \cite{butwick_maternal_2018}. Epidural injections are also used to relieve pain in the chest, abdominal, and lower extremity surgeries and for chronic pain relief \cite{buenaventura2009systematic}. %\cite{wilkinson2012epidural, buenaventura2009systematic}. 
%It is estimated that more than 9 million epidural steroid injections are performed annually in the US \cite{cohen2015single}. %, and approximately 2 million epidurals were performed to relief pain during childbirth in the US in 2015 \cite{butwick_2018}. 
The two most common errors or complications of epidural analgesia are failed epidurals (FE), which are insertions of the needle to a superficial location (that will not produce analgesia and cause the need to perform a re-do), and dural punctures (DP), %\cite{agarwal_complications_2009}, 
occurring in 0.4-6\% of cases \cite{berger_north_1998}, leading to post-dural puncture headaches in 70-80\% of the cases \cite{LEWIS2000238} %\cite{banks_audit_2001}, \cite{LEWIS2000238} 
and subsequently longer hospitalizations \cite{kwak_postdural_2017}, higher chronic pain rates \cite{binyamin2021chronic}, and postpartum depression. The most prominent risk factor for both errors is the experience of the anesthesiologists \cite{task_difficulty_experience_BMI} %\cite{hollister_minimising_2012, task_difficulty_experience_BMI}
who attain competency within 1-90 attempts \cite{konrad_learning_1998}.

% Fig layers
\begin{figure}[!b]
\centerline{\includegraphics[width=8cm]{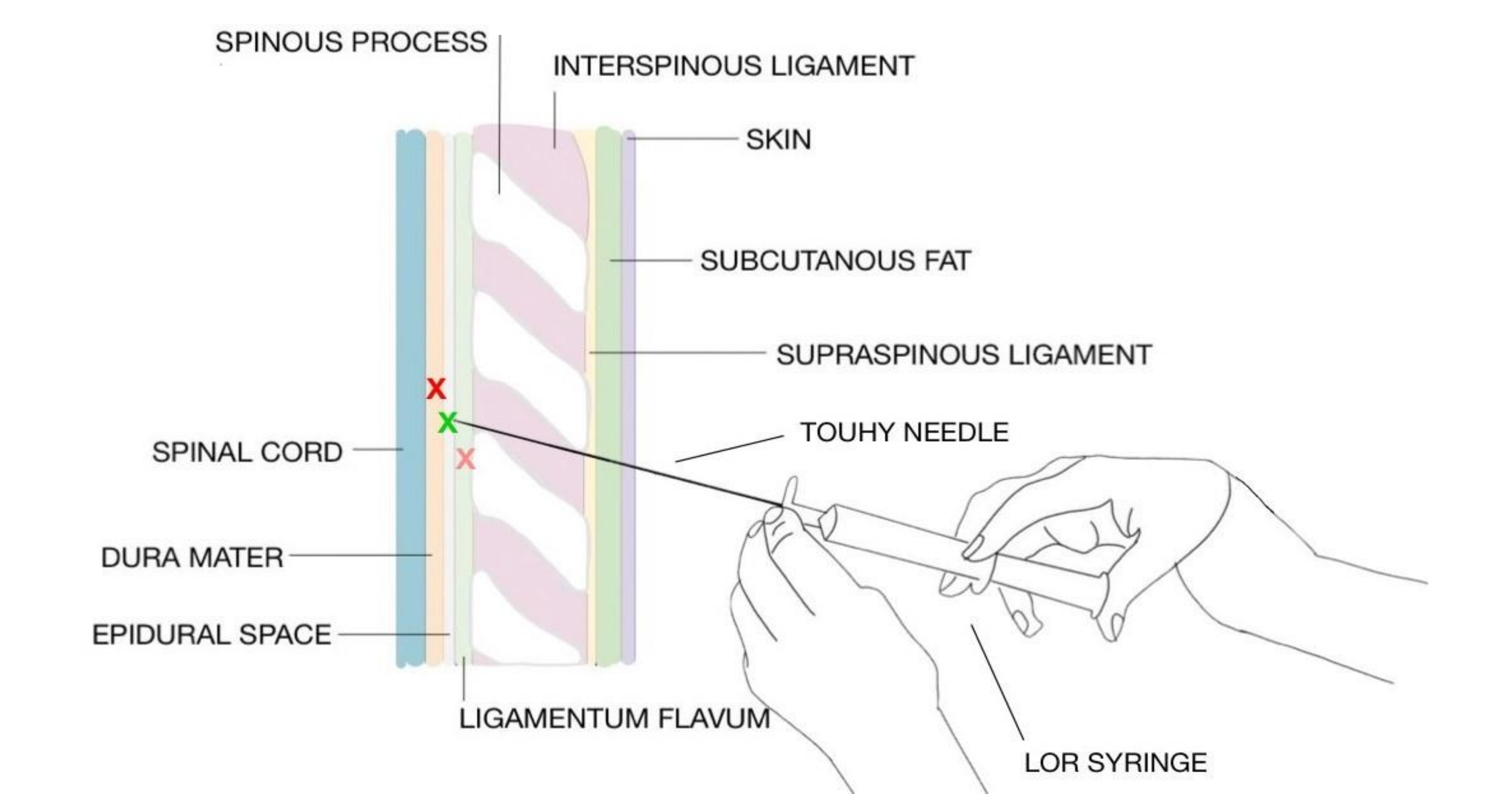}}
\caption{The epidural region. The left and right hands are holding the Touhy needle and LOR syringe, respectively. The Touhy needle is inserted into the epidural space; Such an insertion would be considered a success (green X symbol). The light red X symbol represents a failed epidural (an insertion too superficial, not reaching the epidural space), and the dark red X represents a dural puncture (an insertion too deep, puncturing the dura mater). The difficulty of the task stems from the lack of visual feedback, as well as the thinness of the epidural space, which is encountered after passing through the rigid Ligamentum Flavum, leaving a very small margin for error.}
\label{fig_layers}
\end{figure}

In the epidural anesthesia procedure, medication is inserted into the epidural space, allowing it to block pain signals from traveling from the spine to the brain, hence producing analgesia \cite{TORSKE2000859}. To arrive at the epidural space, the anesthesiologists use an epidural (Touhy) needle, i.e., a hollow needle with a curved tip \cite{pioneers, Capogna2020}. The needle is inserted into the patient's skin in the back area. It proceeds through several tissues located between the skin and the epidural space (Fig. \ref{fig_layers}) while avoiding the spinous process. Once the Touhy needle is ultimately inserted into the epidural space, a catheter may be threaded through the Touhy needle and remain there after the needle has been removed. The anesthetic is then administered using said catheter. The task of stopping the needle immediately after passing the rigid Ligamentum Flavum (i.e., the final tissue preceding the epidural space) is mechanically challenging and requires extensive technical training \cite{konrad_learning_1998}. The thickness of the epidural space in the lumbar region is 5-17 millimeters (depending on patient parameters) \cite{brazil_haptic_2018}, which creates a very small margin for error, and continued movement with the needle will cause an Accidental Dural Puncture (ADP). This is exacerbated by patient variability \cite{task_difficulty_experience_BMI} (e.g., due to obesity or age) affecting the dimensions and stiffness of the layers \cite{brazil_haptic_2018, vaughan_review_2013, lee_mediseus}. 

Since epidural needle insertion is a blind procedure, the "loss of resistance" (LOR) technique helps identify the epidural space as the Touhy needle encounters the various ligaments of the lumbar vertebral column \cite{dogliotti_research_1933, Capogna2020} (Fig. \ref{fig_layers}). This technique is bimanual; the anesthesiologists insert the Touhy needle into the ligaments with one hand and attach a minimal-friction syringe filled with air or saline to the end of the needle with the other hand. Then, they push the needle to advance it and sense the stiffness by applying a force to the plunger with their thumb. The tougher and more fibrous Ligamentum Flavum applies a high resistive force on the needle and the fluid, but once the needle enters the epidural space, the resistance is lost (hence LOR) \cite{LOR_technique, Capogna2020}. This haptic feedback informs the operator of the needle location within the various tissue layers and loss of resistance from potential spaces \cite{LOR_technique}. Since the task is complex and relies on delicate haptic information, several systems assisting epidural needle insertions have been developed, including \cite{li2022admittance}, which presented a novel controller, providing augmented force perception to the operator and newly emerging techniques that assist by guiding the needle to the epidural space, or by helping the operator correctly identify the loss of resistance \cite{Capogna2020}.

Due to the challenges of the task, the procedure is considered the most difficult technique among anesthetic manual tasks to be learned by residents \cite{task_difficulty_tolearn, konrad_learning_1998}. However, even in state-of-the-art training curricula for anesthesiology residents, they do not acquire knowledge and skills in a controlled and safe environment under close supervision. Instead, the training is performed as part of their clinical practice, which is limited by the residents' restricted working hours and, more importantly, jeopardizes patient safety \cite{HoubenKarinW2011Ntsf, Rodriguez-Paz}. 

Medical simulation may solve the trade-off between allowing for high-quality skill acquisition and unharmed patient care. Integrating simulations into resident training programs allows for certain skills to be acquired before meeting the first patient in a clinical setting \cite{Rodriguez-Paz}. Furthermore, practicing a range of manual skills in a virtual environment can be standardized and adjusted to create the desired settings. Residents can repeatedly train at any time until mastering the task \cite{CVC_simulation} while maintaining patient safety \cite{HoubenKarinW2011Ntsf}. Simulation-based training is beneficial as an addition to traditional training methods in simulated hypoxemia and hypotension scenarios \cite{PARKChristineS2010AoCI} and in simulated central venous catheter (CVC) insertions \cite{CVC_simulation}. In the latter, the effects were shown to diminish over time, and, thus, training was suggested to be maintained by repetitive simulation practice. Medical simulation was also shown to be effective in simulating needle insertions and lumbar punctures -- commercially available haptic devices have been successfully utilized for needle insertion simulators in the past \cite{coles2010effectiveness, li2021enhanced}. Another application of medical simulation is the assessment of trainee level \cite{HoubenKarinW2011Ntsf}, which may potentially eliminate the effect of subjectivity in resident-level assessment in many manual tasks. 

In the development of any type of simulator, it is essential to thoroughly investigate its validity. A simulator is considered valid if it accurately reflects the task it is designed to simulate with regard to the specific learning goals and the target population \cite{harris_framework_2020}. There are three primary types of validity to consider in simulation design: face validity, content validity, and construct validity. Face validity is the subjective view users have of how realistic a simulation is, and it can be evaluated using subjective reports, e.g., questionnaires about the resemblance level of each part of the simulator to the real-life environment \cite{harris_framework_2020, manoharan2012design, gavazzi2011face, lee_mediseus}. When examining face validity, it is important to take into consideration the users’ ability to respond to questionnaires and rely more on responses given by users with rich experience in the real-life task \cite{harris_framework_2020, sankaranarayanan2016face, stanton1998learning}. Content validity, another subjective assessment of simulators, examines the usefulness of a simulation in its original purpose \cite{yaghmaie2003content}, e.g., as a training or an assessment tool, as viewed by expert users \cite{yaghmaie2003content, hung2011face}.  

Construct validity is a more objective assessment of simulators, compared to face and content validity; it determines whether the simulator can differentiate between real-life experts and novices. As such, it is considered one of the most important aspects of simulation evaluation \cite{gavazzi2011face}. Construct validity testifies to the simulator’s ability to provide an accurate representation of the real task; a simulation with good construct validity should be able to accurately reflect the differences in performance among individuals, such as between real-life novices and experts, and also within individuals as they develop over time. This would show that the simulation is consistent with the real-world principles \cite{harris_framework_2020, lee_mediseus}. Testing construct validity can be performed through expert versus novice comparisons \cite{harris_framework_2020, harris_exploring_2021, bright_face_2012, manoharan2012design, gavazzi2011face, lee_mediseus}. Its assessment is more objective than face and content validity as it can be tested in a more neutral and impersonal manner. 

Several epidural simulators have been designed since the early 1980s, and they consist of two types: manikins \cite{vaughan_review_2013, broom_evaluation_2018, capogna2018objective} and robotic \cite{lee_mediseus, leleve2019designing, manoharan2012design, vaughan2012haptic, brazil_haptic_2018}. Manikin simulators allow the palpation of the vertebral column, which assists the anesthesiologists in identifying the correct initial needle insertion site \cite{broom_evaluation_2018, capogna2018objective}. However, they do not support any patient variability nor allow modification of the training difficulty, geometry, or mechanical resistance of the tissue with trainee progress \cite{vaughan_review_2013}. 

Robotic simulators use haptic devices to render resistive forces \cite{vaughan_review_2013} that emulate the forces applied to the Touhy needle when encountering the different tissues in the epidural region \cite{leleve2019designing}. Such simulators enable representation of various body types \cite{vaughan_review_2013, lee_mediseus, manoharan2012design}. Additionally, kinematic data, such as injection trajectories and velocities, can be recorded to provide information on the trainees’ progress and improvement with practice. Examples of previous haptic epidural simulators include an electrical haptic interface coupled with a pneumatic cylinder \cite{leleve2019designing} and a motor with a pulley-cable mechanism for transmitting forces in the needle insertion direction \cite{manoharan2012design}. In another simulator \cite{vaughan2012haptic}, the force data was obtained in a needle insertion trial on a porcine cadaver and used to recreate the feeling of epidural insertion. 

Robotic epidural simulators are effective and promising \cite{vaughan_review_2013}, but there is still much room for further development. Previous simulators were uni-manual, although this task is performed bimanually in real-life. They simulated only the forces applied to the Touhy needle without considering those applied to the LOR syringe, an important tool for perceiving the environment stiffness \cite{dogliotti_research_1933, LOR_technique}. Additionally, using the kinematic data acquired by both haptic devices -- and, more specifically, the data regarding probing movements performed using the LOR syringe -- allows for a deeper analysis of strategies, which can assist in the understanding of better training approaches \cite{davidor2023using}. This cannot be done in a uni-manual configuration. Furthermore, a smart training program based on sensorimotor learning principles \cite{herzfeld_motor_2014, dhawale_role_2017} has never been introduced into these robotic simulators. Although robotic simulators today can facilitate patient-based variability, a training program incorporating variability and other motor learning principles has never been designed and implemented.

Some robotic simulators, such as the Mediseus \cite{lee_mediseus}, provide, in addition to the haptic feedback, visual feedback as to the needle's location within the patient's body during the procedure. This may be especially beneficial at the beginning of the training to build a mental map between the experienced forces and the penetration into the different layers. However, even though it may assist the user, visual feedback does not exist in the real procedure and, therefore, may be unnecessary and impede transfer to real procedures. Therefore, in the current study, we chose to develop and assess our robotic simulator without the addition of such visual feedback.

We developed a novel simulator that incorporates the haptic feedback, which is received in both hands, using two haptic devices. By doing so, we provide the user with haptic information not only to emulate the forces applied to the Touhy needle but also to the LOR syringe. Additionally, we implemented the simulation with patient weight variability. This was done to account for the different patient body types that affect the task in the real-life procedure and to set the ground for testing the added value of motor variability and its ability to enhance learning in this task in the future. We present the simulator design and the results of an assessment study. To test face and content validity, we collected the answers to visual analog scale (VAS) questionnaires testing the simulator's resemblance to the real-life task and its suitability as a training or assessment tool. To test construct validity, we compared the success rates of anesthesiologists with different competency levels. We also examined error rates and sizes across participant levels. Further, we studied the movement control strategies that participants employed by measuring their velocities in the different layers of the epidural region and utilization of the probing tool -- the number of probing movements participants performed with the LOR syringe, their depth, and rate. 

\section{Methods}
\subsection{Simulator Design}
\begin{figure}[b]
\centerline{\includegraphics[width=9cm]{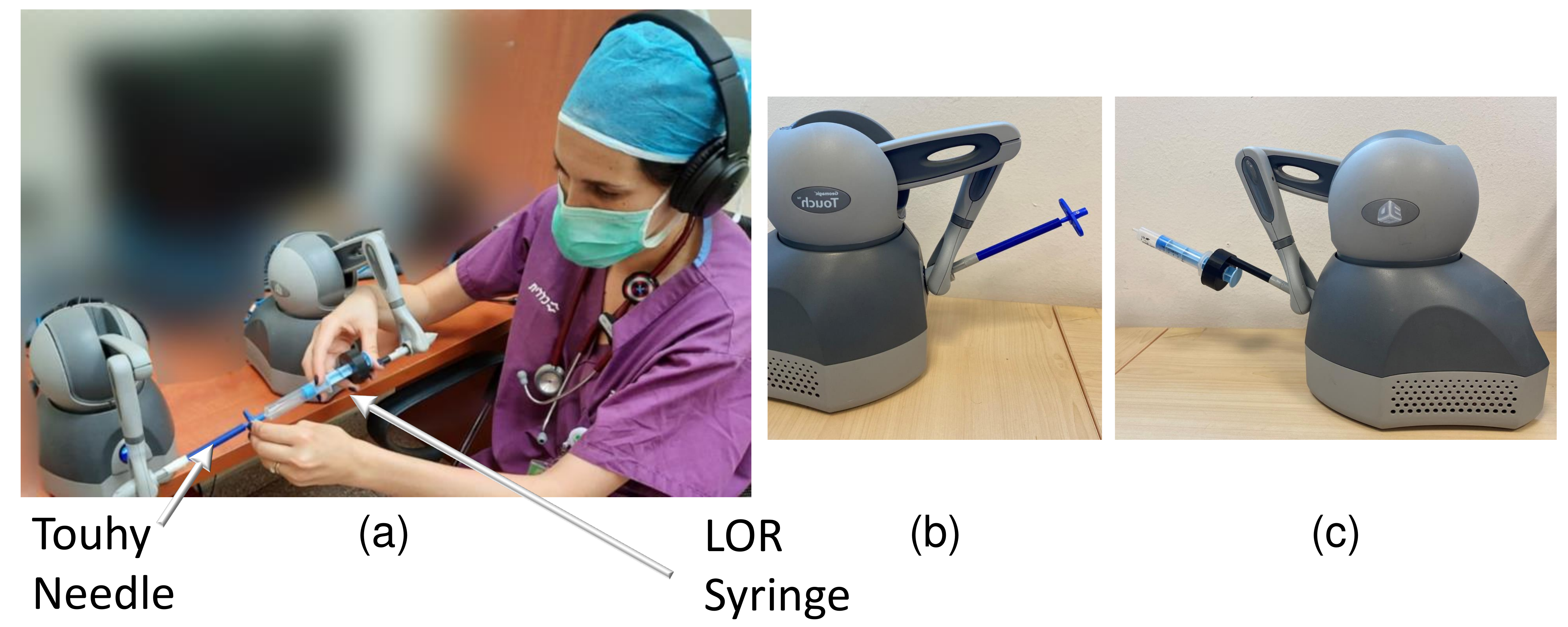}}
\caption{(a) Bimanual epidural simulator, (b) one haptic device is connected to a Touhy needle, and another (c) to an LOR syringe. Both are connected using custom-made 3D-printed adapters. Note that the simulator provides haptic information only.}
\label{fig_sim}
\end{figure}

% hardware paragraph:
We designed a bimanual simulator using two 'Phantom Omni' haptic devices (Fig. \ref{fig_sim}a). One haptic device is mounted by a Touhy needle (Fig. \ref{fig_sim}b), and the other is connected to an LOR syringe (Fig. \ref{fig_sim}c). Both the Touhy needle and the LOR syringe were connected to the haptic devices using 3D-printed custom-made adapters using an 'Ultimaker S5' printer. 

% Fig simulator

% software paragraph:
To render resistive forces, we used the 'Chai-3D' C++ simulation framework (the code will be released upon publication). We used Visual Studio 2013, on a 'Dell Inspiron 13' laptop. The use of a laptop, in addition to the two haptic devices (which are light-weighted), allowed for transporting the simulator to different hospitals or medical conferences to reach anesthesiologist participants population. 

\begin{table*}[ht!]
    \caption{Tissue Force Parameters (Touhy Needle, Average Patient Body Mass)}
    \centering
  %  \begin{tabular}{p{1cm} p{1cm} p{1.25cm} p{1.25cm} p{1.25cm} p{1.25cm} p{3.5cm}}
      \begin{tabular}{|c| c| c| c| c| c| c|}
    \hline
    \textbf{Tissue} & \textbf{Stage} & \textbf{a$_0$ [N]} & \textbf{a$_1$ [N/mm]} & \textbf{a$_2$ [N/mm$^2$]} & \textbf{a$_3$ [N/mm$^3$]} & \textbf{$\Delta $ d range [mm]} \\
    \hline
    Skin & BP & 0.0075 & 0.0037 & $-$0.0015 & 0.0008 & $0<\Delta d<13.92$ \\
    \hline
    Fat & AP & 1.9212 & 0.1437 & $-$0.1682 & 0 & $13.92<\Delta d<17.15$ \\
    \hline
    \multirow{2}{10em}{Supraspinous Ligament} & BP & 0.628 & 0.2637 & 0.0343 & 0 & $17.15<\Delta d<19.37$ \\
    & AP & 1.3855 & $-$0.7174 & 0.0923 & 0 & $19.37<\Delta d<20$ \\
    \hline
    \multirow{2}{10em}{Interspinous Ligament} & BP & 1.4021 & 0.3054 & 0 & 0 & $20<\Delta d<23.18$ \\
    & AP & 2.3761 & 0 & 0 & 0 & $23.18<\Delta d<41.18$ \\
    \hline
    \multirow{2}{10em}{Ligamentum Flavum} & BP & 2.3761 & 0.4783 & $-$0.0186 & 0 & $41.18<\Delta d<44.79$ \\
    & AP & 3.861 & $-$0.0539 & $-$0.0375 & 0 & $44.79<\Delta d<48.38$ \\
    \hline
    Epidural Space & -- & 0 & 0 & 0 & 0 & $48.38<\Delta d<56.98$ \\
    \hline
    \end{tabular}
    \label{table1}
\end{table*}

\begin{figure}[b!]
\centerline{\includegraphics[width=9cm]{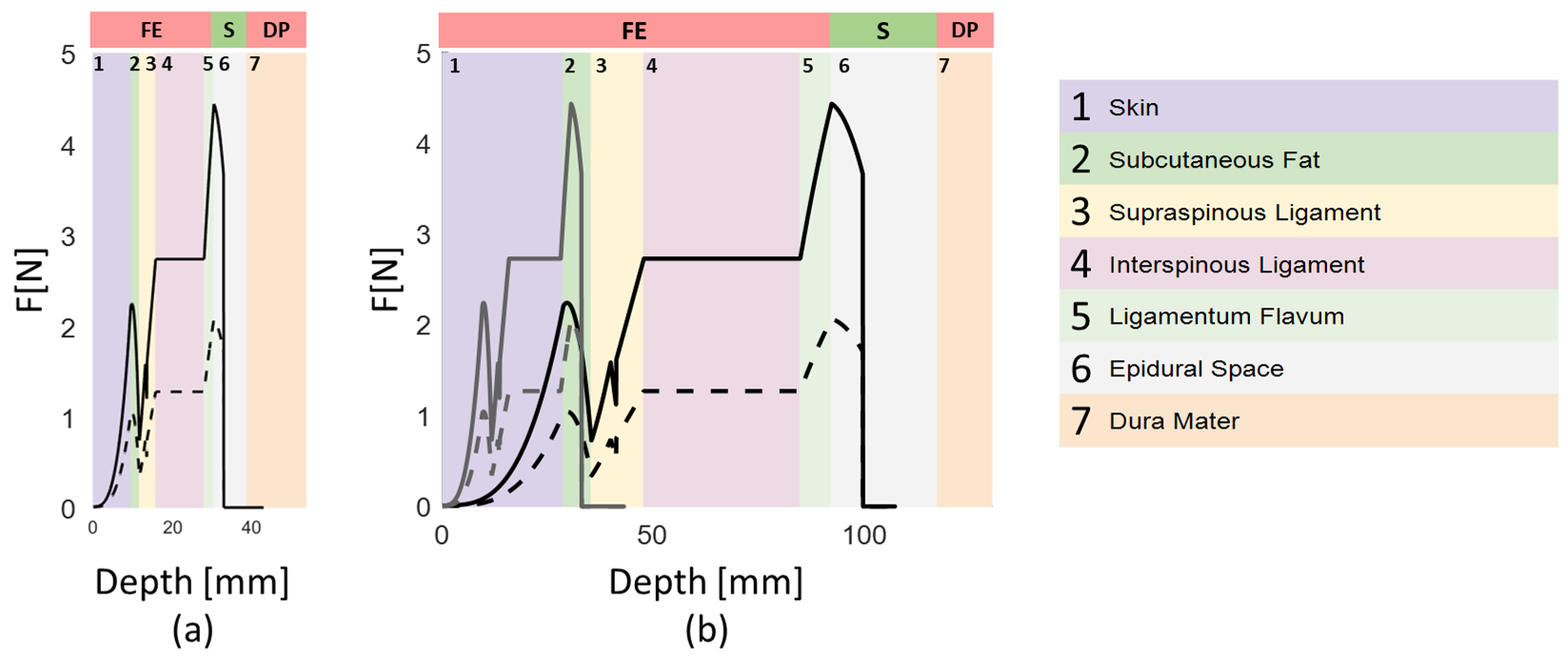}}
\caption{Forces exerted by haptic devices for two different patient body masses: (a) 55 kg and (b) 115 kg (black lines). For reference, the 55 kg patient model is also presented in (b) (gray lines). The ordinate is the exerted force in Newtons. The dashed lines represent the forces exerted by the haptic device connected to the Touhy needle, and continuous lines represent the forces exerted by the haptic device connected to the LOR syringe. The abscissa is the needle penetration depth in millimeters, with results (top boxes) corresponding to the chosen injection site: failed epidural (FE), success (S), and dural puncture (DP). The background colors and numbers represent the layers in the epidural region, as depicted in the color map.}
\label{fig_forces}
\end{figure}

% simulated environment paragraph: 
We relied on the model proposed in \cite{brazil_haptic_2018} to render resistive forces as the trainees cross the different layers in the epidural region (Fig. \ref{fig_forces}). According to this model, each tissue can be emulated by creating a different non-linear spring -- polynomials with different orders and parameters mapping the instantaneous penetration depth of the needle into the tissue layer to a force applied by the haptic devices according to:

\begin{equation}
F=a_0+a_1\Delta d+a_2\Delta d^2+a_3\Delta d^3,\label{eq_polynomials}
\end{equation}

\noindent where $\Delta d$ represents the needle penetration depth into the tissue layer (in mm), and $a_0, a_1, a_2$ and $a_3$ are constants that differ between the different tissue layers and regions within each layer, and their values are presented in Table \ref{table1}. 

This model considers each tissue in the epidural region and subdivides the needle insertion forces in most tissue layers into two regions: before and after the puncture of the tissue layer (BP and AP, respectively). The forces applied before the puncture (BP) result from the stiffness forces that are affected by the tissue elasticity before its perforation. The forces applied after the puncture (AP) can be modeled as a sum of the cutting force, which is the force required to cut through the tissue, and the frictional force, caused by the needle shaft’s contact within the tissue \cite{brazil_haptic_2018}. Because the Subcutaneous Fat is encountered right after the puncturing of the skin, the skin layer includes only the BP stage, and the fat comprises only the AP stage. The epidural space is not a layer that can be punctured, but rather a space to be entered into, and hence does not have BP and AP regions and generally has zero resistance. Note that although the skin is commonly regarded as a tissue that requires a very high cutting force, the Ligamentum Flavum has been shown in several force models (including \cite{brazil_haptic_2018} and \cite{li2021enhanced, esterer2017hybrid}) to require an even higher cutting force.

% Fig forces

\subsubsection{Empirical adjustment of the model}
To account for the limitations of our haptic devices (3DSystems Touch), such as high friction and a limited force range, we implemented some adjustments to the force model proposed in \cite{brazil_haptic_2018}. The purpose of these adjustments was to amend the model such that the forces would be as realistic as possible by verifying it with an expert user. We iteratively changed the parameters with the help of an expert anesthesiologist (author YB) until he found the model appropriate.
As a result of this process, the parameters in Table \ref{table1} are adjusted compared to those proposed in \cite{brazil_haptic_2018}, such that the parameters in \cite{brazil_haptic_2018} for the haptic device connected to the Touhy needle, and for a patient of an average body mass of 71 kg, are multiplied by a scaling factor of 0.3. Additionally, we set the epidural space's force to zero (unlike the value of 2.5 N, proposed in \cite{brazil_haptic_2018}) to intensify the sense of loss of resistance. To render the forces applied by the second haptic device, which is connected to the LOR syringe, we used a scaling factor of 2 ($F_{LOR}=2 \cdot F_{Touhy}$). %The relation between the forces applied to the Touhy needle and those applied to the LOR syringe is described in equation \ref{eq_2.1factor}: 
We opted not to incorporate the deflection model proposed in \cite{brazil_haptic_2018} for simplicity and since this is the first prototype of the simulator.

%\begin{equation}
%F_{LOR}=2.1 \cdot F_{Touhy}, \label{eq_2.1factor}
%\end{equation}

%\noindent where $F_{LOR}$ is the force applied to the LOR syringe, and $F_{Touhy}$ is the force applied to the Touhy needle (both in Newtons).

\subsubsection{Introducing patient weight variability}
To add patient weight-related variability, we adopted an adjustment to the model of the epidural regions of female patients with variable body masses proposed in \cite{brazil_haptic_2018}. We added variability in layer thickness and stiffness according to:

\begin{equation}
T_t=A_t\left(\frac{\sqrt{A_w(B_m/A_m)/\pi}}{R_w} \right)^3\label{eq_weight}
\end{equation}

\noindent where $T_t$ is the tissue thickness, $A_t$ is the average tissue thickness (according to $\Delta d$ ranges in Table \ref{table1}). $A_w=574.94 $ $ cm^2$ is the average female waist area, $R_w=13.53 $ $ cm$ is the average female waist radius, $B_m$ is the relevant body mass in kg, and $A_m=71 $ $ kg$ is the average body mass for a female patient aged between 20-29 years; all according to \cite{brazil_haptic_2018}. We used the average patient weight of 71 kg in some of the trials and three different body masses of 55, 85, and 115 kg when we wanted to add patient weight variability (see details in the 'experimental procedures' section). These figures were inserted into Eq. \ref{eq_weight}, with the different values of $B_m$ affecting the tissue thickness. 

To account for the patient weight effect on tissue stiffness, we divided $\Delta d$ by $T_t/A_t$. This scaling yielded changes in the stiffness and thickness of each layer according to the patient body mass. However, the value of the rendered force in the layer was unchanged, as depicted in Fig. \ref{fig_forces}b, where two patients of different body masses (55 kg in gray in 115 kg in black) are shown. In this example, the challenge is highlighted -- the epidural space of the patient with the lower body mass is more superficial than that of the patient with the higher body mass and is at a depth equal to that of the higher body mass patient's Subcutaneous Fat.   

\subsection{Experimental Procedures}
The experiment included two sessions of data collection: the first ($N_1=7$) took place in Soroka Medical Center. The second session ($N_2=15$) was conducted during the 2021 Israel Society of Anesthesiologists annual conference (Tel Aviv, Nov 2021). We set up a booth at the conference and invited the attendees to participate. In total, 22 anesthesiologists from several hospitals in Israel, with different levels of experience, participated in our experiment. The participants were volunteers and had no previous connection with the authors. They all signed an informed consent form, as stipulated by the human participants' research committee approval, and were not compensated for their participation. To assess their skill level, we asked participants to fill out questionnaires in which we inquired about their years of experience, an estimated number of epidural injections they had performed thus far, and whether they were residents or attendings. 

\begin{table}[b!]
\caption{Anesthesiologist Level Assignment System}
\centering
\begin{tabular}{|cl|l|l|l|}
\hline
\multicolumn{2}{|c|}{\textbf{\begin{tabular}[c]{@{}c@{}}Categorical\\ Level\end{tabular}}} &
  \multicolumn{1}{c|}{\textbf{\begin{tabular}[c]{@{}c@{}}Years of \\ Experience\end{tabular}}} &
  \multicolumn{1}{c|}{\textbf{\begin{tabular}[c]{@{}c@{}}Estimated \\ Number\\  of Epidurals\end{tabular}}} &
  \multicolumn{1}{c|}{\textbf{Position}} \\ \hline
\multicolumn{2}{|c|}{\textbf{1}} & 0 to 1  & 0 to 50    & Resident  \\ \hline
\multicolumn{2}{|c|}{\textbf{2}} & 1+ to 3 & 50+ to 300 &           \\ \hline
\multicolumn{2}{|c|}{\textbf{3}} & Over 3  & Over 300   & Attending \\ \hline
\end{tabular}
\label{table2}
\end{table}

Participants sat next to a table on which the two haptic devices were placed. They were instructed to hold the Touhy needle (connected to one haptic device) with their non-dominant hand and the LOR syringe (connected to the second haptic device) with their dominant hand, similar to the technique used in real procedures. During the experiment, the participants wore headphones with active noise cancelling effect (Bose QC35) to eliminate auditory cues, including noises produced by the haptic devices, and more specifically, a noise produced when arriving at the epidural space and losing resistance. Such noises are not present in the real-life procedure and might affect performance, and hence should be eliminated. The participants received no auditory or visual information, as this simulator version provides solely haptic feedback. 

During the first data collection session, the experiment consisted of 12 test trials. Since this protocol lacked a period of acclimatization to the simulator, we decided to add three familiarization trials at the beginning, such that in the second data collection session, the experiment consisted of 15 trials: three familiarization trials followed by 12 test trials. In each trial, the participants were instructed to perform a virtual needle insertion using the simulator and verbally indicate when they reached the selected injection site (i.e., where they perceived the epidural space was, and where they would insert the epidural catheter). Participants were told the patient's body mass at the beginning of each trial, as such information is available in the real-life procedure. During the familiarization trials, participants were allowed to familiarize themselves with the virtual environment and were provided feedback on the trial result. A ‘success’ would be identifying the epidural space correctly. For a patient with an average body mass, as shown in Table \ref{table1}, a trial would be considered successful if the participant identified the epidural space in the range of $48.38 < \Delta d_f < 56.98 $ mm (where $\Delta d_f$ is the final needle penetration depth, where the participant chooses to halt and provide a verbal cue). A ‘failed epidural’ would be an undershoot ($\Delta d_f < 48.38$ mm), which in real-life would result in the need to perform a re-do, and a ‘dural puncture’ would be an overshoot ($\Delta d_f > 56.98$ mm). During the test trials, participants received no feedback as to the trial result. 

The forces exerted by the haptic devices in the familiarization trials were designed to emulate the epidural region of a female patient with an average weight of 71 kg (Table \ref{table1}). In the test trials, the exerted forces emulated three different patient body masses; 55, 85, and 115 kg. Each weight appeared four times, and the order of the trials was mixed into four blocks of the three patient weights.

\subsection{Data Analysis}
% data collection paragraph:
\subsubsection{Data recording}
\begin{figure}[b!]
\centerline{\includegraphics[width=9cm]{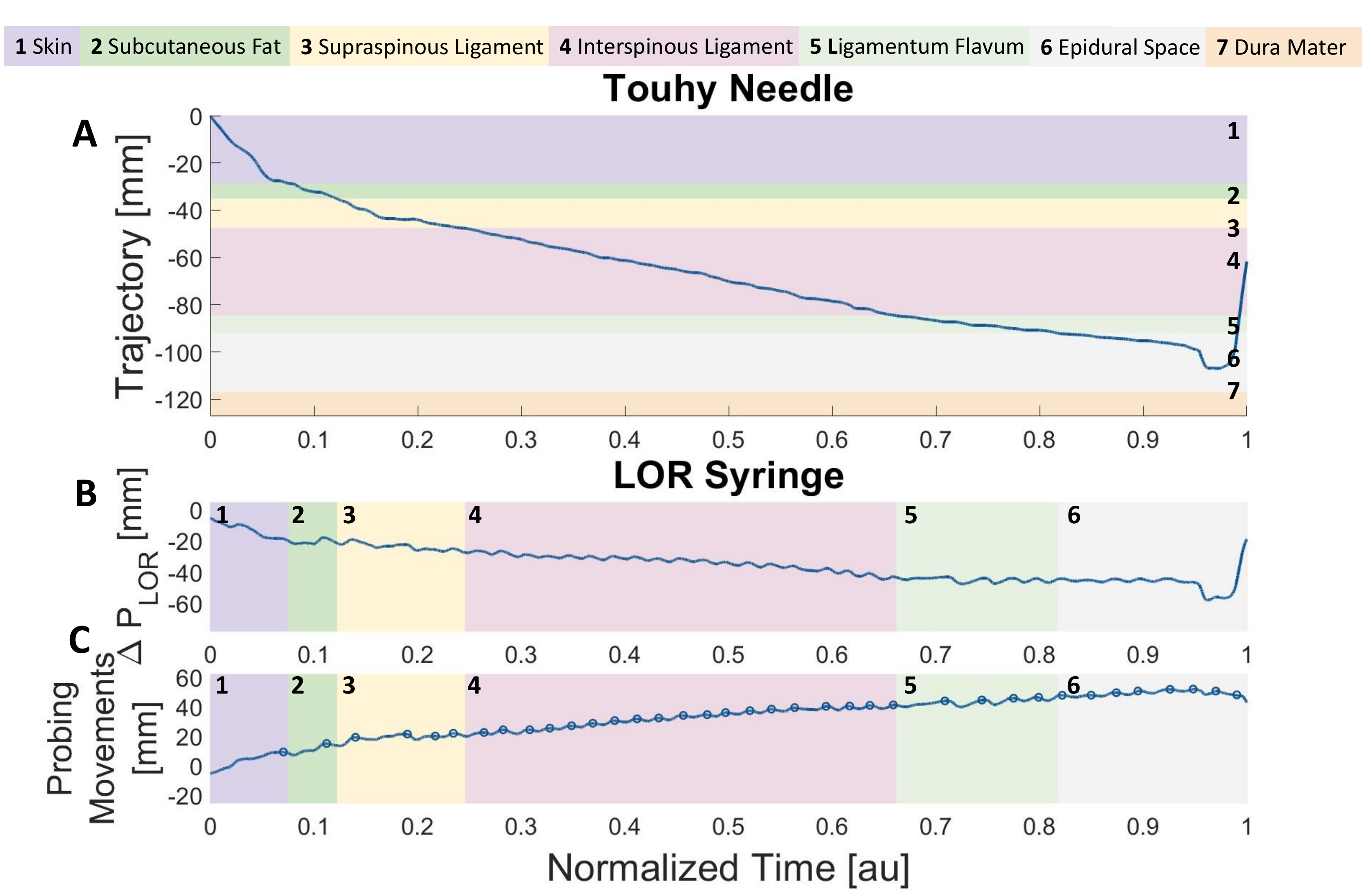}}
\caption{Probing movement obtaining: (A) The trajectory of the haptic device connected to the Touhy needle. (B) The trajectory recorded by the haptic device connected to the LOR syringe, where the two haptic devices' 'zero' points are calibrated. (C) The adjusted trajectory (according to eq. \ref{eq_adjust_P_LOR}), and detected probing movements. The background colors and numbers represent different tissue layers in the epidural region, as depicted in the color map.}
\label{fig_probe_calc}
\end{figure}

We recorded the kinematic data (insertion trajectory, velocity, and final injection site) of both haptic devices (each of them separately) at approximately 1 kHz. This allowed online error feedback (which was provided to participants in the familiarization trials, see more in the 'experimental procedures' section) and offline data analysis for trainee evaluation and progress monitoring. Since the data was recorded from both haptic devices, we were able to examine the kinematics of the Touhy needle and the LOR syringe separately (Fig. \ref{fig_probe_calc}, \ref{fig_trajectory}). Therefore, we could observe not only the proceeding through the different tissues in the epidural region (which is performed using the Touhy needle) but also examine the probing movements performed with the LOR syringe, which are used to perceive the environment stiffness. We did not use the familiarization trials for the data analyses but rather the test trials only. To eliminate differences in familiarization between the two data collection sessions, we used only the final nine trials of the data collected during the first session. 

% Level Assignment %
Anesthesiologists' levels (novices N=6, intermediates N=9, and experts N=7) were determined according to a tailored level assignment system (Table \ref{table2}), which took into consideration the anesthesiologists' years of experience, an estimated number of epidural injections they had performed prior to participating in the study, and whether they were residents or attendings. For each of the three categories, we determined the categorical level per participant (the level could be assigned as either one, two, or three). Next, to achieve the overall level of the anesthesiologist, we averaged and rounded the levels.

\subsubsection{Performance and validity analysis}
To study the performance in the task, i.e., to quantify the rate of the three result types: failed epidural, success, and dural puncture, we recorded the final injection site chosen by the participant in each trial. Then, we examined the result for each trial and the rate of each result per participant throughout the test trials. We compared the failed epidural, success, and dural puncture rates of participants in the three different levels. The success rate allows for assessing the trainee's level of competency when performing the procedure in the virtual environment. When examining the trainee's errors, we can learn more about their strategies. For instance, a higher rate of failed epidurals could imply excessive caution, as opposed to a higher rate of dural punctures, which could indicate recklessness.

To assess the simulator's construct validity, we aimed to test users' performances in the simulated environment. The reasoning behind that was that demonstrating that real-life experts perform better with the simulator compared to novices will show good construct validity. To do so, we examined the success rates (the metric best suited to assess performance level) of participants in different experience levels. In addition, we used the final injection sites to calculate the error size in millimeters in each trial. This was defined as the distance of the chosen injection site from the correct epidural space location. A successful trial would yield an error of 0 mm, a failed epidural would yield a negative error, and a dural puncture -- a positive one. We took the absolute value of the error, so as not to allow failed epidurals and dural punctures to cancel out one another. We compared errors in absolute value and anesthesiologist levels.
    
To evaluate the simulator’s face and content validity, we asked participants to respond to questionnaires regarding the level of resemblance of each layer to the tissue in real-life. There were three additional questions regarding the compatibility of the simulator as a training and assessment tool, and the participants' overall impression of the simulator. To collect the responses, we used the 100 mm visual analog scale (VAS). The VAS is a continuous scale, 100 mm long, which allows participants to mark an estimated response to a question upon it \cite{crichton2001visual}. When using this scale for estimating realism (compared to the real-life environment), '0' represents 'completely unrealistic' and '100' represents 'completely realistic'. When using it for testing the simulator's suitability as a training or assessment tool, '0' represents 'completely unsuitable' and '100' represents 'completely suitable'. %Based on results observed in the validation of previous epidural simulators \cite{pedersen2017loss, lee_mediseus, broom_evaluation_2018, raj_simple_2013}, we determined a threshold of 70\% to be the landmark of a 'good' score.

 Four of the participants in this study did not fill out the questionnaires, as they had so little experience with the real-life environment, they did not feel confident to compare between the real-life environment and the virtual one (hence the smaller sample size compared to the number of participants in the study). We divided the remaining 18 participants in our analysis into two groups: anesthesiologists who had performed less than 500 epidurals prior to participating in this experiment ('Inexperienced', N=11), and anesthesiologists who had performed 500 or more epidurals prior to this experiment ('Experienced', N=7). We chose this method for dividing the participants since we aimed to examine the degree to which participants perceive the virtual environment as similar to the real-life one. Therefore, it was crucial to separate between the responses of individuals that have a rich experience with the real-life environment and those who do not. Note that among the participants in the inexperienced group, some had performed only a few epidurals (under 20). For each question and group, we calculated the mean scores and their 95\% confidence intervals.

% Fig probe calc

\subsubsection{Probing kinematics analysis}
We quantified how the participants used the LOR syringe. To do so, we recorded kinematic data from both haptic devices separately. We used only the data recorded in the part that emulated the patient back -- starting from the point where the penetration depth of the device connected to the Touhy needle ($\Delta d$) was zero. Then, to obtain the probing movements with the LOR syringe, we subtracted the location of the device connected to the Touhy needle from the location of the device connected to the LOR syringe (eq. \ref{eq_adjust_P_LOR}). This allowed us to separate between the probing movements and the Touhy needle proceeding through the layers in the epidural region. Due to the LOR syringe length (10.5-15 cm, depending on how deep inside the syringe the plunger is), there is an inherent distance between the two haptic devices. To allow for calibrated 'zero' points, we subtracted this length from the trajectory of the LOR syringe as well. To obtain the exact length per trial, we took the location of the haptic device connected to the LOR syringe where $\Delta d=0$.

\begin{equation}
P_{adj.}=\Delta P_{LOR}-P_{Touhy},  P_{Touhy}>0 \label{eq_adjust_P_LOR}
\end{equation}

\noindent where $P_{adj.}$ is the adjusted trajectory, $\Delta P_{LOR}$ is the location of the haptic device connected to the LOR syringe (after the 'zero' point calibration), and $P_{Touhy}$ is the location of the haptic device connected to the Touhy needle; all units are provided in millimeters. 

The main purpose of this adjustment was to separate the peaks that occurred due to LOR probing movements from those stemming from sudden movements with the Touhy needle. To find the number of probing movements, we used the 'findpeaks' function in Matlab on the adjusted trajectory; to assess the penetration depth of each probing movement, we measured each peak's height in millimeters. To attain the probing movement rate, we measured the time difference between each pair of consecutive peaks, and used its inverse. To assess the number of probing movements and the mean penetration depth in each trial, we used the number of probing movements enumerated in each trial, and for each probing movement, its depth. Then, we calculated the mean penetration depth across all probing movements in the trial. To obtain the probing movement rate per trial, we calculated the mean of the inverse time differences throughout the entire trial. Subsequently, to test them against anesthesiologist level or trial result, we calculated the mean number of probing movements, mean penetration depth, and mean rate across all relevant trials (according to the factor we wanted to test it against). 

Additionally, we examined the location of each probing movement within the different layers of the epidural region. We normalized the number of probing movements in each layer by dividing it by the thickness of the layer in millimeters. We chose this normalization of the number of probing movements since the presence inside a tissue for a larger distance would provide more opportunities to probe. Thus, probing more in a thicker tissue would not necessarily imply that there was uncertainty in this tissue. 

We also calculated the mean velocity of the haptic device connected to the Touhy needle in each layer in the epidural region for each participant. 

% Fig trajectory
\begin{figure*}[ht!]
\centerline{\includegraphics[width=\textwidth]{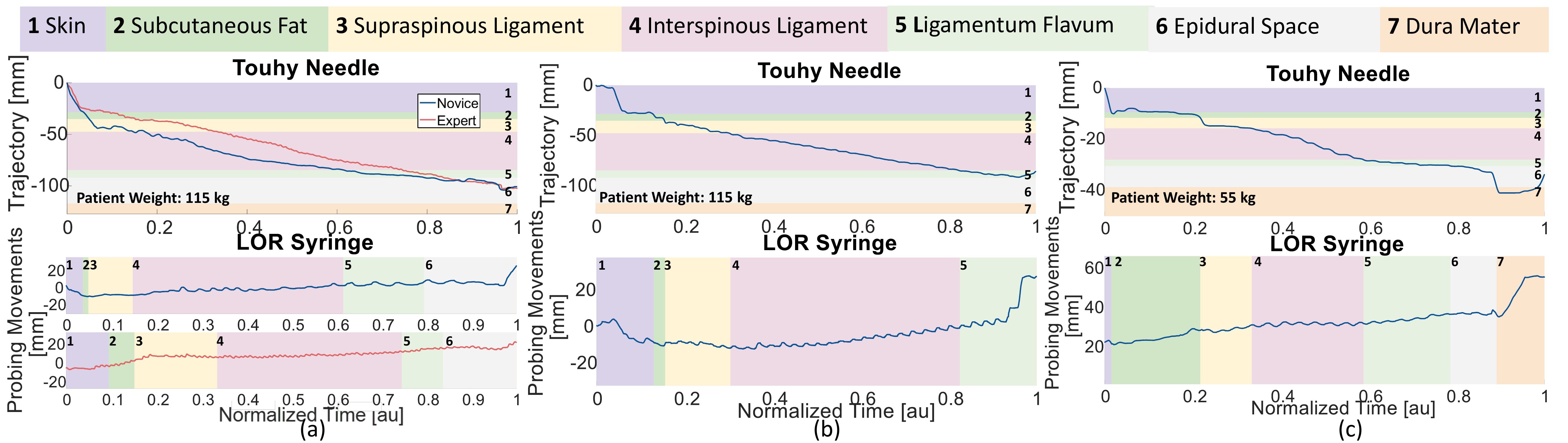}}
\caption{Kinematic data recorded by the two haptic devices. In all panels, the top boxes represent the participants' trajectories with the Touhy needle %(the abscissa is the normalized time (arbitrary units), and the ordinate is the needle penetration depth into the patient back (in millimeters))
and the bottom boxes represent probing movements with the LOR syringe (see a detailed description of the calculation method in the 'Data Analysis' section). Different background colors and numbers represent different tissue layers in the epidural region, as depicted in the color map for tissue layers. The patient weight correlated to the trial is denoted in the top boxes. (a) Trajectories and probing movements as performed by a novice and an expert. % (the blue line represents a novice, and the red line represents an expert). 
Both participants succeeded in this trial. (b) Trajectories and probing movements of a novice participant in a trial that resulted in a failed epidural. (c) Trajectories and probing movements of an intermediate-level anesthesiologist in a trial that resulted in a dural puncture. }
\label{fig_trajectory}
\end{figure*}

\subsubsection{Statistical Analysis}
\begin{figure}[hb!]
\centerline{\includegraphics[width=9cm]{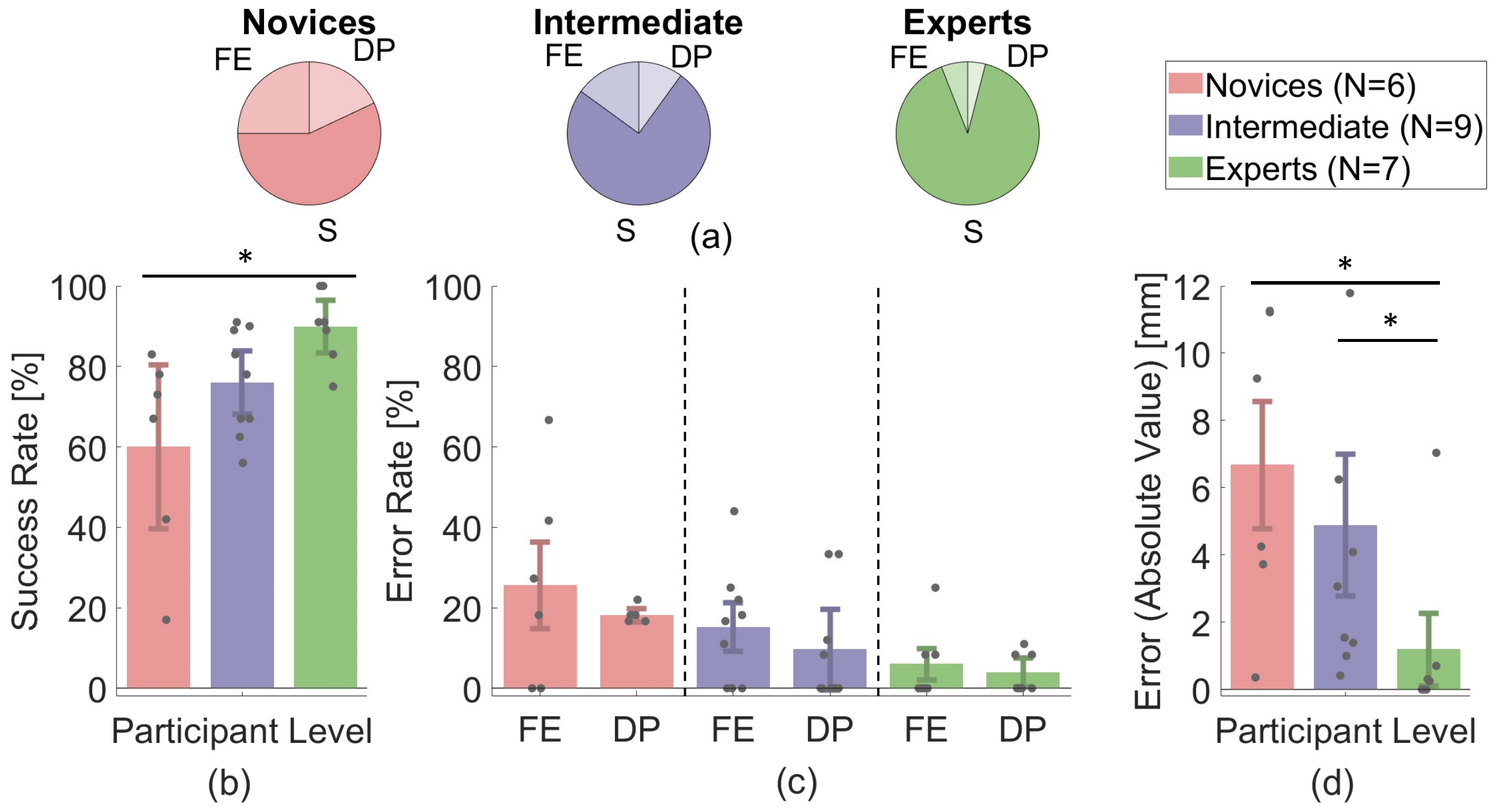}}
\caption{Result rates as a function of participant level. %The different colors represent the different anesthesiologist levels: pink  represents novices (level 1), purple  represents intermediate level anesthesiologists (level 2) and green represents experts (level 3). 
(a) Pie charts of mean result rates for all three results (failed epidural, success, and dural puncture); each chart represents a different anesthesiologist level. (b) Colored bars represent success rate. The gray points represent each participant's mean success rate, the error bars represent the mean confidence intervals, and the asterisks indicate a statistically significant difference ($p<0.05$). (c) Colored bars represent error rates, i.e., failed epidural (FE) and dural puncture (DP). Dashed black vertical lines represent the separation between the different participant levels. The gray points represent the rate each participant achieved for each of the two error types. (d) Colored bars represent the mean absolute value of the error in millimeters. The gray points represent the absolute value errors of each participant.}
\label{fig_result_rates}
\end{figure}

In our statistical analysis, we compared several contrasts: result rates, error size, probing movement enumeration, probing movement penetration depth, and probing movement rate. None of the metrics were distributed normally, and hence, we chose non-parametric statistical methods. For each metric, we calculated the mean and the non-parametric bootstrap 95\% confidence intervals. We used the Kruskal-Wallis test to determine whether there were statistically significant differences between the success rates (Fig. \ref{fig_result_rates}b), error sizes (Fig. \ref{fig_result_rates}d), number of probing movements, probing movement penetration depths, and probing movement rates (Fig. \ref{probes_levels}a-c) of the different anesthesiologist levels. In these models, the three categorical independent variables were the anesthesiologist levels (novices, intermediate, and experts), and the dependent variables were the success rates each participant achieved, the absolute value errors calculated for each participant, the mean number of probing movements per trial, the average probing movement depth, and the probing movement rate. When we found a significant effect of expertise level, we performed post-hoc pairwise comparisons with the Wilcoxon rank-sum test and used the Bonferroni correction for multiple comparisons.

The analysis of probing movement numbers, their penetration depth, and their rate against the trial result (Fig. \ref{probes_levels}d-f) would ideally be analyzed with a repeated-measures model. However, since the metrics were not normally distributed, and there were many missing data points (only 6 of the 22 participants had all three data points), we could not fit models that accounted for the repeated measures, and chose instead to analyze these results using the Kruskal-Wallis test. 

To analyze the differences in error rates within participant levels, we used the Wilcoxon signed-rank test and compared failed epidural rates and dural puncture rates in each of the three anesthesiologist levels separately (Fig. \ref{fig_result_rates}c). 

% Fig result rates

\section{Results}
To help explain how we calculated the metrics introduced in this section, we first present a visual demonstration of the kinematic data we recorded to provide an intuition about needle insertion trajectories of successful and unsuccessful trials, along with different LOR probing strategies performed by participants with different levels of experience. 

In Fig. \ref{fig_trajectory}, we display several examples of trajectories performed by the hand holding the Touhy needle alongside probing movements with the LOR syringe. Fig. \ref{fig_trajectory}a depicts the trajectories and probing movements of two participants, a novice and an expert, in a trial that was successful for both participants. %It is noticeable that both participants did not perform a failed epidural or a dural puncture -- both their trajectories reached the gray area, which represents the epidural space. 
The LOR probing trajectories demonstrate that the participants had different strategies; The expert participant (red) probed more often and more superficially than the novice (blue). This result is aligned with a trend we observed for most of the participants. An example of a trial that resulted in a failed epidural performed by a novice participant is shown in Fig. \ref{fig_trajectory}b. The needle insertion is halted within the green area, representing the Ligamentum Flavum, which indicates that the participant did not arrive at the epidural space. An example of a trial that resulted in a dural puncture performed by an intermediate-level participant is shown in Fig. \ref{fig_trajectory}c, where the Touhy needle crossed through the epidural space (gray area) and arrived at the Dura Mater (orange area).
 
To assess the construct validity of our simulator, we examined the effect of anesthesiologist level on success rate in the virtual task. In Fig.~\ref{fig_result_rates}a, the distribution of the different result types in all the trials of novice, intermediate, and expert participants are depicted, highlighting that novices make the most errors, followed by intermediate and, finally, the expert participants. Fig. \ref{fig_result_rates}b depicts that the success rate indeed depends on the participant's level of expertise, supported by a statistically significant effect of expertise (Kruskal-Wallis yielded $\chi^2=8.41$, $p=0.0149$, $df=2$). Post-hoc analysis revealed a significant difference between novices (level 1) and experts (level 3), ($p_{1-3}=0.0129$, $p_{1-2}=0.5240$, $p_{2-3}=0.1147$).

We also studied the rates of the two error types (failed epidural and dural puncture) separately for the three anesthesiologist levels (Fig. \ref{fig_result_rates}c). %We observed hints to different trends between the different levels: In novices (level 1) and experts (level 3), the failed epidural rate appeared to be higher than the dural puncture rate, and in intermediate level anesthesiologists (level 2), the dural puncture rate appeared to be higher than the failed epidural rate.%
We observed the same trends in all participant levels; the failed epidural rate appears to be higher than the dural puncture rate. However, these results were not statistically significant (Wilcoxon signed-rank tests yielded $p_1=0.5411$, $p_2=0.3711$, $p_3=0.75$), which prevents us from drawing concrete conclusions. 

  % Fig errors
%\begin{figure}[!b]
%\centerline{\includegraphics[width=4cm]{abs_err.png}}
 %   \caption{Errors corresponding to participant level. The abscissa represents participant level. %(left to right: pink bars represent novices (level 1), purple bars represent intermediate level anesthesiologists (level 2) and green bars represent experts (level 3)). 
  %  The ordinate represents the mean absolute value of the error in millimeters. The black bars represent the mean confidence intervals. The gray points represent the  absolute value errors of each participant.}
%\label{fig_errors}
%\end{figure}

To further analyze errors, we assessed the effect of the anesthesiologist level on error size (measured as the absolute distance value of the chosen injection site from the correct epidural space location). The results depicted in Fig. \ref{fig_result_rates}d demonstrate an inverse relationship between participant level and error; Novices' errors appeared to be larger in size than those of intermediate-level anesthesiologists and experts. This result was supported by statistical significance (Kruskal-Wallis yielded $\chi^2=8.38$, $p=0.0152$, $df=2$). Post-hoc analysis revealed a significant difference between novices (level 1) and experts (level 3), and between intermediate-level anesthesiologists (level 2) and experts (level 3) ($p_{1-3}=0.0241$, $p_{2-3}=0.0436$, $p_{1-2}=0.8850$).

\begin{figure}[b!]
\centerline{\includegraphics[width=9cm]{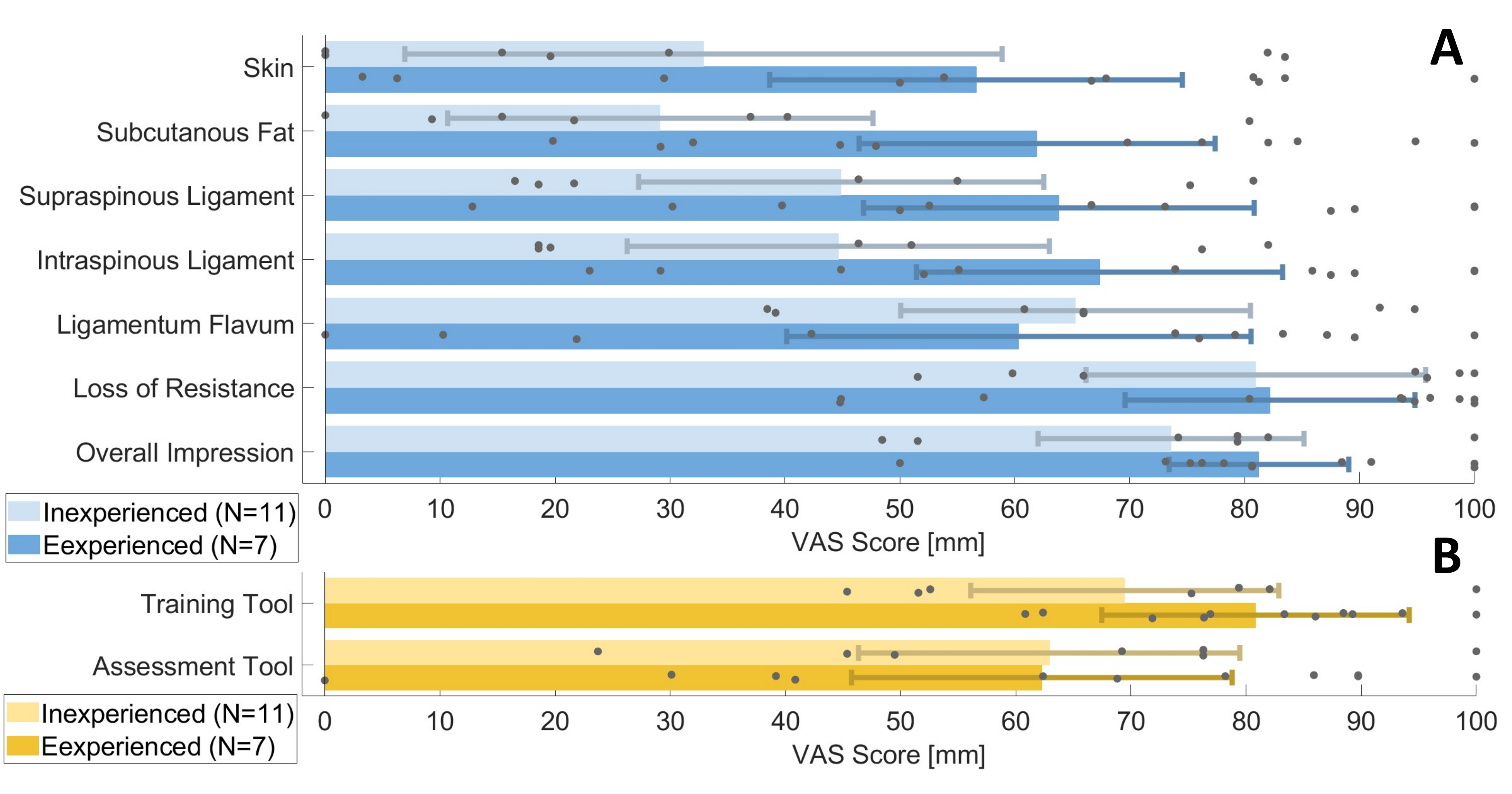}}
\caption{Visual analog scale questionnaire responses evaluated by two groups: inexperienced and experienced. (A) Face validity: blue bars represent the mean VAS score calculated for each of the different layers in the epidural region and their similarity to the real-life tissue and the overall impression. (B) Content validity: yellow bars represent mean VAS scores calculated for the compatibility of the simulator as a training and assessment tool. In both (A) and (B), the error bars represent the mean confidence intervals, and the gray points represent the VAS scores given by each participant in the relevant category.}
%anesthesiologists who had performed less than 500 epidurals prior to their experience with the simulator (N=5, light colored bars), and anesthesiologists who had 500 or more epidurals prior to their experience with the simulator (N=6, dark colored bars). (a) The blue bars represent the mean VAS score calculated for each of the different layers in the epidural region and their similarity to the real-life tissue (light blue is for group 'Under 500', and dark blue is for 'Over 500'). (b)  Yellow bars represent mean VAS scores calculated for overall impression, and compatibility of the simulator as training and assessment tools (light yellow is for group 'Under 500', and dark yellow is for 'Over 500'). }
\label{fig_VAS}
\end{figure}

To assess face and content validity, participants' responses to VAS questionnaires were evaluated (Fig. \ref{fig_VAS}) and presented separately for two groups ('Inexperienced' and 'Experienced'). The responses' means were calculated for each question (for each tissue layer and for three questions evaluating the simulator) and for each group of participants. In most layers of the epidural region (all layers but the Ligamentum Flavum, Fig. \ref{fig_VAS}a), the responses of the experienced group were higher than or equal to those of the inexperienced group. This was also the case for the evaluation of the overall simulator impression and its suitability as a training tool (Fig. \ref{fig_VAS}b). The overall impression mean score of the experienced group, as well as this group's mean score for the loss of resistance and suitability as a training tool, were relatively high (82.22 $\pm$ 14 mm, 81.24 $\pm$ 22.2 mm, and 80.82 $\pm$ 12.4 mm, respectively).

% Fig VAS

% Fig probes X levels
\begin{figure}[hb!]
\centerline{\includegraphics[width=9cm]{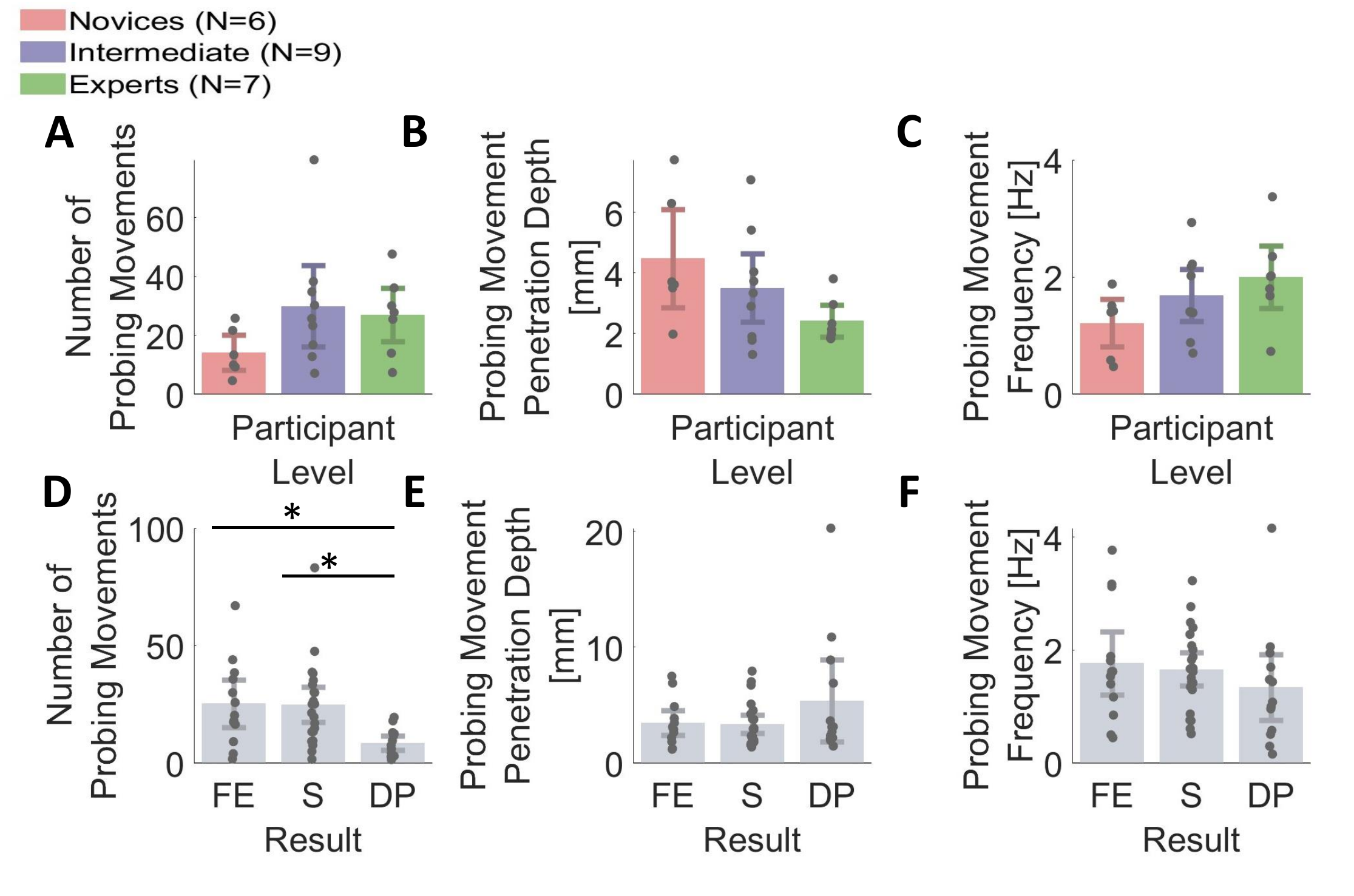}}
\caption{Probing movements. The abscissa in (A)-(C) is the participant level, and in (D)-(F), it is the trial outcome. %(left to right: pink bars represent novices, purple bars represent intermediate level anesthesiologists, and green bars represent experts) 
The ordinate in (A),(D) is the mean number of probing movements performed in each trial, in (B),(E) - the mean penetration depth of these  probing movements, and in (C),(F) - the probing movement frequency. The gray points represent the mean values for each participant. The error bars are the mean confidence intervals, and the asterisks indicate statistical significance ($p<0.05$).}
%number of probing movements performed by each participant in (a), the mean probing movement penetration depth for each participant in (b) and the probing movement rate for each participant in (c). }
\label{probes_levels}
\end{figure}

% Fig probes X results
%\begin{figure}[t]
%\centerline{\includegraphics[width=9cm]{probes_result_withFreq.png}}
%\caption{Probing movements for different trial outcomes. The abscissa is the trial result (left to right: failed epidural (FE), success (S), and dural puncture (DP)) and the ordinate is (a) the mean number of probing movements performed in each trial, (b) the mean penetration depth of the probing movements performed in each trial and (c) the probing movement frequency. The gray points represent the mean values for each participant. The black bars represent the mean confidence intervals, and the asterisks indicate a statistically significant difference ($p<0.05$). }
%number of probing movements performed by each participant in (a), the mean probing movement penetration depth for each participant in (b) and the probing movement rate for each participant in (c). The black bars represent the mean confidence intervals
%\label{probes_results}
%\end{figure}

To study probing strategies, we assessed the effect of anesthesiologist level on the number of probing movements and their penetration depths and rate (Fig. \ref{probes_levels}a-c). It appeared that experts and intermediate-level anesthesiologists used a larger number of probing movements compared to novices, and these movements were more superficial and fast, although all three results were not significant (Kruskal-Wallis tests ($df=2$) yielded $\chi^2_{num}=4.27$, $p_{num}=0.1182$, $\chi^2_{depth}=3.316$ $p_{depth}=0.1911$, $\chi^2_{rate}=3.83$, $p_{rate}=0.1471$). 

Additionally, we assessed the effect of each trial's result on the number of probing movements in the trial. Fig. \ref{probes_levels}d demonstrates that utilizing a small number of probing movements led to a higher chance of causing a dural puncture (Kruskal-Wallis yielded $\chi^2=12.03$, $p=0.0024$, $df=2$). Post-hoc analysis revealed a significant difference between the number of probing movements in trials that resulted in dural punctures and trials that resulted in both failed epidurals and successful trials ($p_{FE-DP}=0.0114$, $p_{S-DP}=0.0037$). No difference was found between the mean number of probing movements in trials that led to failed epidurals and successful trials ($p_{FE-S}=1$). Examining the probing movements' depths and rate (Fig. \ref{probes_levels}e-f) indicated that the movements used in successful trials and trials that resulted in failed epidurals were often more superficial and fast than the probing movements in trials that resulted in dural punctures, although this result was not significant (Kruskal-Wallis ($df=2$) yielded $\chi^2_{depth}=0.85$, $p_{depth}=0.6527$, $\chi^2_{rate}=2.23$, $p_{rate}=0.3273$). 

\begin{figure}[t]
\centerline{\includegraphics[width=9cm]{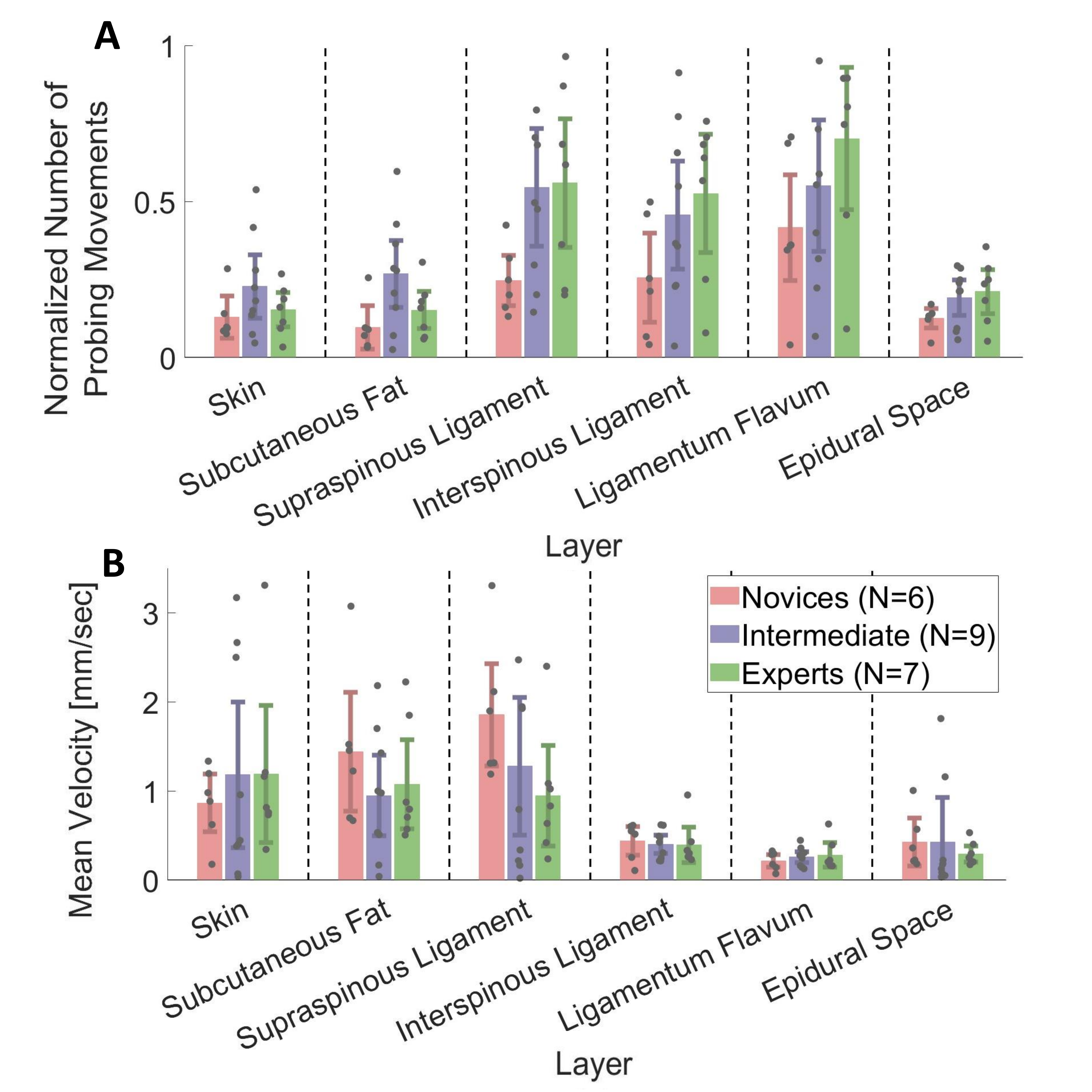}}
\caption{Location analysis. The abscissa is the layer of the epidural region, and the ordinate in (A) is the mean normalized number of probing movements per trial. Colored bars represent the mean number of probing movements performed in each trial for each layer in the epidural region, divided by the layer thickness in millimeters. The ordinate in (B) is the mean velocity. %(left to right: Skin, Subcutaneous Fat, Supraspinous Ligament, Interspinous Ligament, Ligamentum Flavum, epidural space). 
The different colors represent the different anesthesiologist levels. %: pink bars represent novices (level 1), purple bars represent intermediate level anesthesiologists (level 2) and green bars represent experts (level 3). 
Dashed black vertical lines represent the separation between the layers. The gray points represent the mean values performed by each participant in the relevant layer, and the error bars represent the mean confidence intervals.}
\label{probes_layers}
\end{figure}

We also conducted an exploratory analysis of the normalized number of probing movements performed (Fig. \ref{probes_layers}a) and the mean velocity (Fig. \ref{probes_layers}b) in each of the layers in the epidural region. Accordingly, we present the analysis results without any statistical analysis. We did not have predictions about the probing strategies for the different layers, and the current study is under powered due to the large number of layers combined with the small number of data points at each anesthesiologist level. Nevertheless, we present these results to be used as a pilot for future studies. 

%Studying the layer in the epidural region in which probes were performed (Fig. \ref{probes_layers}a) indicates that all participants probed the greatest number of probes in the Interspinous Ligament, in which the forces are constant. There is a relation between participant level and the number of probes in most layers (4 out of 6 layers): it appeared that experts probed more than novices. 
Studying the layer in the epidural region in which probing movements were performed (Fig. \ref{probes_layers}a) indicated that all participants probed the greatest number of probing movements in the three layers preceding the epidural space: the Supraspinous Ligament, Interspinous Ligament, and Ligamentum Flavum. There is a relationship between participant level and probing movements' number in most layers (4 out of 6 layers); it appeared that experts and intermediate-level anesthesiologists probed more than novices in these layers. 

We also examined the mean velocity of the Touhy needle in the different layers of the epidural region; the results depicted in Fig. \ref{probes_layers}b demonstrate that participants decreased their velocity in the Interspinous Ligament (and continued with a decreased velocity afterward). %Furthermore, experts and intermediate level anesthesiologists seem to have decreased their velocity in the Supraspinous Ligament, which is the layer preceding the Interspinous Ligament. 

% Fig probes X layers

\section{Discussion}
In this study, we designed and assessed a novel bimanual simulator for epidural analgesia, using two haptic devices and custom-made adapters to the instruments used in the real-life procedure (a Touhy needle and an LOR syringe). The simulator is novel in the sense that it is bimanual, allowing for the emulation of the forces applied not only to the Touhy needle, as was done thus far, but to the LOR syringe as well. %Additionally, the modeling of 
We used a force model based on the one proposed in \cite{brazil_haptic_2018}, and scaled it following an iterative process of adjustment to the experience of an expert anesthesiologist, which allowed us to render resistive forces, considering the haptic devices' limitations and striving for maximum realism. 
We modeled multiple patient body weights that allowed for motor variability which has the potential to enhance motor learning \cite{dhawale_role_2017}. 
We conducted a study with 22 anesthesiologists of different experience levels. Our relatively small sample size primarily results from the specialized nature of the participant population (anesthesiologists), which limited the feasibility of obtaining a more extensive sample. We found that the simulator was able to distinguish between real-life novices and experts, demonstrating good construct validity. Face and content validity, as assessed by the VAS questionnaire responses, varied between the different questions. In some, the more experienced users' satisfaction level was low, and we suggest some interpretations for this result in this section. In others, the mean scores provided by experienced users were relatively high, suggesting good face and content validity, though validating the simulator accordingly remained difficult, due to several limitations presented in this section. Analysis of the kinematic data demonstrated differences in strategies between anesthesiologists of different experience levels, and implied a relationship between strategies and outcomes in the virtual task. 

The relation between anesthesiologist level and their success rate in the virtual environment validates the simulator's construct validity according to \cite{harris_framework_2020}, as it demonstrates that the simulator is able to distinguish between real-life novices and experts. The statistically significant difference between novices and experts supports this claim. Testing construct validity in this manner has been adopted in the past by several groups \cite{harris_framework_2020, harris_exploring_2021, bright_face_2012, bright2014visual}, including \cite{harris_exploring_2021}, who compared the performance of novice and expert golfers in a golf VR simulation, and \cite{bright_face_2012}, who compared experts and novice surgeons in minimally invasive surgical simulations. Bright et al. \cite{bright2014visual} observed significant differences between experts and novices in a surgical simulator. Additional to differences in performance level, they also revealed differences in control strategies, which provided a novel method of assessing construct validity in medical simulators. In this section, we also address differences in strategies observed between novices and experts and suggest methods of utilizing these characteristics to enhance training with our simulator.    

In the evaluation of the VAS questionnaire responses, we chose to put more emphasis on the experienced group, as these are more experienced users we can rely on to compare the virtual and real-life environments. Further, evaluating face and content validity using expert opinions is commonly used in the medical simulator field and beyond \cite{harris_framework_2020, sankaranarayanan2016face, yaghmaie2003content}. In most layers and in the overall impression (Fig. \ref{fig_VAS}a), the experienced group's mean scores were relatively high, and the 'Loss of Resistance', which is the most important haptic aspect required to succeed in the task (as it informs the anesthesiologist of reaching the epidural space), received the highest score in the layer category.

These results suggest good face validity; however, due to several reasons, we are still limited in our ability to validate the simulator. First, it is difficult to interpret the subjective ratings without comparison to other systems (e.g., a uni-manual epidural simulator). If the literature would propose an official threshold for what is a ’good’ VAS score to determine face validity, we could compare to such threshold. However, such gold standard does not exist, and the absolute scores are difficult to interpret and be used to determine the validity of the simulator. Further investigation and comparison to other simulators will be necessary to validate the simulator in the future. 

Additionally, there are several limitations in the simulator design, that we would expect to be better reflected in participant reports: these limitations include insights that we received from several expert anesthesiologists (including the author YB), that the Ligamentum Flavum has a unique texture, which is not well-emulated in this simulator. Other limitations include the lack of the deflection model proposed in \cite{brazil_haptic_2018}, and the absence of visual information, which highly affects face validity \cite{harris_framework_2020}. We aim to address all these issues in future versions of the simulator. The fact that these limitations did not affect the VAS scores as much as we would expect testifies to the weakness of these subjective evaluations. 

In this version of the simulator, we chose not to incorporate visual information because we posit that the haptic aspect of the task is the most important. This is because the needle insertion task is a blind procedure (the needle is inserted into the patient's skin, and the anesthesiologist does not receive any visual information as to its location). We believe that this choice has contributed to the construct validity of our simulator, but adding visual information might improve user satisfaction and affect face validity. Hence, in future versions of the simulator, we will consider adding visual information that aims to emulate the real-life environment by simulating a delivery suite in which a patient is sitting on a hospital bed. Such visual information should not include any feedback regarding the needle location within the epidural region since this visual information is not provided to the anesthesiologist in the real-life procedure.

The simulator was perceived by experienced users as a better tool for training than assessment (Fig. \ref{fig_VAS}b). We designed this simulator for training, and hence, this result implies good content validity. However, it is possible that in their answers, the participants did not consider all relevant factors. For example, if they would consider cost and training time, they might lower the scores of the suitability of the simulator as a training tool. Such perspective was demonstrated in \cite{pedersen2017loss} and \cite{raj_simple_2013}, where inserting the needle into a banana was preferred over other simulators that were rated more realistic because of its cost efficiency. In addition, our analysis of the performance and strategies revealed differences between different experience levels of anesthesiologists. This means that in contrast to the answers of participants, the simulator can be used as an objective assessment tool. Both aspects highlight the challenges of validating the simulator based on subjective participant reports.

%The simulator was perceived by experienced users as a better tool for training compared to an assessment tool (Fig. \ref{fig_VAS}b). This result agrees with our original design for the simulator as a training tool alone, implying good content validity. However, it is possible that participants did not consider all relevant factors when responding to these questions; for instance, if participants would consider the cost of the system and the time that would need to be spent on training, they might lower the scores given to the suitability of the simulator as a training tool. Additionally, after analyzing the results of this study and observing differences in strategies between different level anesthesiologists, we believe that examining user strategies in depth can be used as an objective assessment tool for anesthesiologists in different stages of training. Validating the simulator based on these participant reports proves challenging due to both regards.

Error rates allow us to study strategies in terms of caution versus recklessness. It appears that all participants erred on the side of safety. This result can imply a level of insecurity; some participants (especially novices) are more likely to fear an over-shoot, thus resulting in mistakenly halting the needle insertion too superficially. For others, this result demonstrates the comprehension of how potentially harmful accidental dural punctures are, explaining why they made such efforts to avoid them, occasionally risking a result of failed epidural. Another possible interpretation for this result is the fact that participants knew their actions were being examined, and hence deployed a higher level of caution.

%Error rates allow us to study strategies in terms of carefulness versus recklessness. Novices and experts both erred to the side of safety. For novices, we argue that this result implies a level of insecurity. Novices are more likely to fear of an over-shoot, thus resulting in mistakenly halting the needle insertion too superficially. For experts, we claim that this result demonstrates their comprehension of how potentially harmful accidental dural punctures are, explaining why experts make such efforts to avoid them, occasionally taking the risk of resulting in a failed epidural. The increased amount of dural punctures, compared to failed epidurals, shown in intermediate level anesthesiologists suggests, according to our hypothesis, that their confidence level is higher than that of novices, however their comprehension of the potential risks, as well as their competency level, is still not as good as that of experts. Hence, they appear to carry out more overshoots and less undershoots, demonstrating a lower level of carefulness. The rate of accidental dural punctures was also examined in the real-life procedures by Orbach-Zinger et al. \cite{orbach-zinger_anesthesiologists_2021}, and showed the same trend portrayed here: intermediate level anesthesiologists performed more dural punctures than both novices and experts.

Measurement of error sizes allows us to examine how far the chosen injection site was from the correct epidural space. The result shown in Fig. \ref{fig_result_rates}d reinforces the result shown in Fig. \ref{fig_result_rates}a-c, as it demonstrates that not only were novices' errors more prevalent than those of intermediate-level anesthesiologists and experts, they were also more prominent. 

Probing movement enumeration, depth, and rate evaluation showed that using less probing movements can lead to a higher risk for dural punctures, and that experts probed more than novices. The use of less probing movements was often accompanied by a larger depth and slower rate; however, this was not supported by our statistical analysis. We argue that experts and intermediate-level anesthesiologists (that were shown to be more successful and perform fewer dural punctures according to Fig. \ref{fig_result_rates}) probe more than novices since they know that utilizing the probing tool can assist in perceiving the environment stiffness and help prevent dural punctures (as shown in Fig. \ref{probes_levels}d). Hence, they choose to use it more often. However, additional studies are needed to ascertain these preliminary observations.

Probing movement location analysis was evaluated in an exploratory manner without statistical analysis. Accordingly, we discuss the results cautiously. Fig. \ref{probes_layers}a indicated that all participants probed the most in the three layers preceding the epidural space. This does not align entirely with our original hypothesis, which would suggest that participants probe the most in the Ligamentum Flavum preceding the epidural space. Instead, it appears that participants began their extensive probing (indicating higher awareness of proximity to the epidural space) earlier. In most layers, experts probed more than novices, which coincides with the result implying that experts utilize the probing tool more than novices. The only exceptions were the Skin and Subcutaneous Fat layers, where experts used the tool less than novices. This supports the previous result, as it demonstrates that the only layers in which experts probed less were those where they were certain they hadn't reached the epidural space yet (due to the fact that the needle insertion had just begun) and hence did not need to make efforts to perceive the environment stiffness.

%Fig. \ref{probes_layers}a insinuated that all participants probed the most in the Interspinous Ligament. This is in contradiction to our original hypothesis, which would suggest that participants probe the most in the Ligamentum Flavum, which precedes the epidural space. However, we hypothesize that the reason for the excessive use of probing in Interspinous Ligament is rooted in the layer's resistance, which is constant with respect to needle penetration depth, and the fact that it is the most thick layer (as depicted in Fig. \ref{fig_forces}). The non-changing resistance reduces the level of certainty and might create an illusion of loss of resistance (which would lead participants to probe the environment in order to perceive its resistance to comprehend their current position). Additionally, the thickness of the layer causes participants to spend more time in this layer, which would provide them with more opportunities to probe the environment. 

Velocity analysis was also evaluated in an exploratory manner, and Fig. \ref{probes_layers}b depicted that all participants had decreased their velocity in the Interspinous Ligament. This result partially coincides with the large number of probing movements presented in the three layers preceding the epidural space in Fig. \ref{probes_layers}a. It appears that although participants had started to probe the environment stiffness in the Supraspinous Ligament (showing a higher level of awareness and preparation towards encountering the epidural space), they still began decreasing their velocity only in the Interspinous Ligament, which is the second of the three mentioned layers. %Furthermore, it appeared in Fig. \ref{probes_layers}b, that experts had decreased their velocity in the preceding layer, the Supraspinous Ligament. This result indicates that experts knew to anticipate the forthcoming layer, and had started preparing for the velocity decrease in advance (which co-aligns with their extensive probing, that began from this layer and onward). 

%This result co-insides with the large number of probes presented in this layer in Fig. \ref{probes_layers}a.Our hypothesis that the Interspinous Ligament confused participants and led them to believe they should act carefully, as they might be approaching loss of resistance, is strengthened by their decrease in velocity, which could also imply uncertainty. Furthermore, it appeared in Fig. \ref{probes_layers}b, that experts had decreased their velocity in the preceding layer, the Supraspinous Ligament. This result indicates that experts knew to anticipate the forthcoming layer, and had started preparing for the velocity decrease in advance. 

The differences in strategies between different level anesthesiologists shown in this section lead us to suggest trainee-specific feedback and instructions in advanced training stages (e.g., for intermediate-level anesthesiologist training). We argue that our simulator is necessary to allow novices to train in a virtual environment before performing their first real-life epidural analgesia procedure, but also that anesthesiologists should continue training with it throughout their residency to maintain skill acquisition level \cite{CVC_simulation}. In advanced training stages, we suggest that trainees be presented with a summary of the strategies they had performed (e.g., how many probing movements they performed, the location in the epidural region in which these probing movements took place, how they adjusted their velocity, etc., with respect to the trial result), and that trainees be instructed with methods to improve, based on expert strategies harvested in this study. This teaching method, which incorporates repeated practice and refined instruction, has been shown as a key step in technical skill acquisition for surgical residents \cite{reznick1993teaching, kopta1971approach}. Furthermore, the trainee-specific approach proposed here is one of the principles that can enhance adult learning \cite{reznick1993teaching}.

%While acknowledging the limitations described in this section, we assessed the face, content, and construct validity of the simulator we designed. By showing good construct validity, which is more objective than face and content validity (since it depends on functional similarity between the simulation and the real task \cite{harris_framework_2020}), we were able to prove the simulator's ability to simulate the task of epidural needle insertion accurately. 

%Our ability to analyze trainee strategies will enable trainee-specific instruction-based training as a form of reinforcement learning that can also be beneficial \cite{dhawale_role_2017} and suitable for adult learning \cite{reznick1993teaching}.
%The integration of the use of the simulator can be an improved training paradigm in epidural analgesia,  

%This work is the first of three planned steps. In the future, we plan to determine a preferred training protocol that will allow for efficient training, by testing several protocols. Some of these will consider the effect of motor variability on training, as well as test explicit versus implicit learning. Once we will establish a training protocol, we plan to test the simulator's ability to improve resident performance in the real-life task of epidural analgesia.

\section{Conclusions}
In this work, we presented the design and assessment of a bimanual simulator for optimizing skill acquisition in epidural analgesia. The simulator we designed comprises two haptic devices connected to the real-life task instruments and emulates the structure and resistance of the epidural region, while allowing for patient weight variability and data acquisition for both online and offline data analysis. We used the simulator to examine the performance of 22 anesthesiologists of different competency levels and demonstrated that it can distinguish between real-life novices and experts, suggesting good construct validity. Experienced users' VAS questionnaire responses were examined to assess face and content validity, and we discussed the limitations in interpretation of the results and challenges in validation of the simulator. Our strategy analysis suggests that experts utilize probing with the LOR syringe (which is a tool for perceiving the environment's resistance) more than novices and that insufficient utilization of said tool yields a higher dural puncture rate.  We argue that the simulator could be used as a tool for training novice anesthesiologists before encountering their first real-life case, and consequently increase patient safety, and that practice with it should be maintained and refined to be trainee-specific over time. Integrating the simulator into the training curricula can improve the training paradigm, compared to the 'see one, do one, teach one' approach implemented today.

\bibliographystyle{IEEEtran}
\bibliography{ref2}

\end{document}